\documentclass[prb,two column,floats,showpacs,amsmath,amssymb]{revtex4}
\usepackage{float}
\usepackage{graphicx}
\usepackage[usenames,dvipsnames]{color}
\usepackage{hyperref}
\usepackage{bbm}

\newcommand{\operator}[1]{\hat{#1}}

\newcommand{\oO}{\operator O}

\newcommand{\oH}{\operator H}

\newcommand{\oS}{\operator S}
\newcommand{\oI}{\operator I}
\newcommand{\orho}{\operator \rho}
\newcommand{\oR}{\operator \rho}
\newcommand{\oP}{\operator P}
\newcommand{\oomega}{\operator \omega}

\newcommand{\mysqrt}[1]{\sqrt{\smash[b]{#1}}}
\newcommand{\tr}[1]{\,\mathrm{tr}[#1]}	% trace
\newcommand{\mean}[1]{\langle #1 \rangle}	% average
\newcommand{\norm}[1]{\Vert #1 \Vert}
\newcommand{\av}[1]{\mean{#1}}
\newcommand{\abs}[1]{\ensuremath{\left\vert #1 \right \vert}} % absolute value
\newcommand{\bra}[1]{\langle #1 |}
\newcommand{\ket}[1]{| #1 \rangle}

\newcommand{\dist}[1]{D_{#1}}	% distinguishability
\newcommand{\ittad}[1]{\Delta_{#1}}	% infinite-times time-averaged distinguishability
\newcommand{\ittadest}[1]{\widetilde{\Delta}_{#1}}	% estimate for ""
\newcommand{\var}{\text{Var}}
\newcommand{\lpineq}{L}

\usepackage{xspace}			%useful for right space after abbreviations

\newcommand{\tad}{TAWD\xspace}
\newcommand{\HI}{\mathrm{HI}}
\newcommand{\ZE}{\mathrm{ZE}}

\newcommand{\bath}{\mathrm{nuc}}

\newcommand{\I}{\mathrm{i}}
\newcommand{\tot}[1]{\text{d}\hspace{-0.8pt}#1\;}

\newcommand{\tone}{\mathbbm{1}}

\newcommand{\ie}{\text{i.\;e.\xspace}}
\newcommand{\eps}{\varepsilon}

\newcommand{\refeq}[1]{Eq.~(\ref{#1})}

\def\clap#1{\hbox to 0pt{\hss#1\hss}}

\def\mathrlap{\mathpalette\mathrlapinternal}
\def\mathclap{\mathpalette\mathclapinternal}

\def\mathrlapinternal#1#2{%
\rlap{$\mathsurround=0pt#1{#2}$}}
\def\mathclapinternal#1#2{%
\clap{$\mathsurround=0pt#1{#2}$}}

\begin{document}

% -------------------------------------------------------------------------------------------------------------------------------------------
% document settings
% -------------------------------------------------------------------------------------------------------------------------------------------

% \setlength{\parindent}{0cm}

% -------------------------------------------------------------------------------------------------------------------------------------------
% beginning of text-body
% -------------------------------------------------------------------------------------------------------------------------------------------

\title{Equilibration in closed quantum systems: Application to spin qubits}

\author{Daniel Hetterich}
\affiliation{Institut f\"{u}r Theoretische Physik und Astrophysik,
Universit\"{a}t W\"urzburg, D-97074 W\"urzburg, Germany}

\author{Moritz Fuchs}
\affiliation{Institut f\"{u}r Theoretische Physik und Astrophysik,
Universit\"{a}t W\"urzburg, D-97074 W\"urzburg, Germany}

\author{Bj\"orn Trauzettel}
\affiliation{Institut f\"{u}r Theoretische Physik und Astrophysik,
Universit\"{a}t W\"urzburg, D-97074 W\"urzburg, Germany}

\date{\today}

\begin{abstract}
We study an ``observable-based'' notion of equilibration and its application to realistic systems like spin qubits in quantum dots. On the basis of the so-called distinguishability, we analytically derive general equilibration bounds, which we relate to the standard deviation of the fluctuations of the corresponding observable. Subsequently, we apply these ideas to the central spin model describing the spin physics in quantum dots. We probe our bounds by analyzing the spin dynamics induced by the hyperfine interaction between the electron spin and the nuclear spins using exact diagonalization. Interestingly, even small numbers of nuclear spins as found in carbon or silicon based quantum dots are sufficient to significantly equilibrate the electron spin.
\end{abstract}

\pacs{03.65.Yz, 05.30.-d,  76.20.+q, 85.35.-p}

\maketitle

% \tableofcontents

\section{Introduction}

% Introduction

The theoretical understanding of the notion of equilibration in closed quantum systems has significantly developed in recent years.\cite{Deutsch1991,Srednicki1994,Short2011,Short2012,Reimann2008,Reimann2012,Rio2014,GarciaPintos2015,Polkovnikov2011} In the absence of a thermal bath and the presence of quantum fluctuations, the classical concepts of the physical and mathematical description of equilibration do in general not work anymore.\cite{Srednicki1999} Therefore, it is of utmost importance to first properly define what we mean by equilibration in closed quantum systems that can even be in a highly non-thermal state. This difficult task is one of the driving forces of the research area of quantum thermodynamics. Useful concepts imply different definitions of equilibration. 
For instance, many authors identify equilibrium with the saturation of the expectation values of certain observables.\cite{Srednicki1999, Reimann2008, Yukalov2011,Eisert2015} These ideas are appealing as they are intuitive and the relevant quantities are measureable. However, as it is argued in Ref.~\onlinecite{Short2011}, this definition is not satisfying because the measurable probabilities of the outcomes of an observable may still be dynamical while its expectation values has saturated. In our article, we start from a more sophisticated concept\cite{Short2011,Short2012} and link it to the above discussed ideas of saturating expectation values. Doing so, we connect abstract definitions of quantum equilibration to a very concrete experimental system where one could potentially see this exciting physics.

% Prominent examples are  devices made from group IV elements, e.g. Si\cite{Abe2010,Wild2012,Steger2013,Zwanenburg2013}, carbon nanotubes\cite{Kuemmeth2008,Churchill2009,Churchill2009a,Steele2009,Monge2011,Chorley2011,Eichler2011,Lai2014,Laird2014}, graphene\cite{Goossens2012,Allen2012,Liu2009,Liu2010a,Stampfer2008,Ponomarenko2008,Schnez2009,Molitor2009,Guttinger2009,Molitor2010,Wang2010,Guttinger2010,Guttinger2011,Fringes2011,Engels2013,Neumann2013,Epping2013,Drogeler2014,Couto2014,Jacobsen2014}, or diamond\cite{Dutt2007,Neumann2008,Togan2010,Togan2011,Doherty2013}, since all of these atoms exhibit nuclear spin-less isotopes.

The system we have in mind is an electron spin confined in quantum dot (QD) that functions as a spin qubit.\cite{Loss1998} In typical host materials, like GaAs,\cite{Hanson2007,Urbaszek2013,Kloeffel2013} silicon\cite{Tyryshkin2003,Sellier2006,Lim2009,Zwanenburg2013,Simmons2007,Kawakami2014} or carbon,\cite{Jelezko2004,Doherty2013,Ponomarenko2008,Guttinger2009a,Goossens2012,Allen2012,Guttinger2012,Liu2009,Liu2010a,Wang2010,Guttinger2011,Fringes2011,Jacobsen2014,Cao2005,Kuemmeth2008,Steele2009,Laird2014} the electron spin is coupled to many nuclear spins by the hyperfine interaction.\cite{Schliemann2003,Coish2009} The number of nuclear spins that matters for the electron spin depends on the size of the QD, i.e. the confinement, and the natural abundance of nuclear-spin carrying isotopes of the host material. In practice, this number can range between very few nuclear spins (in silicon- or carbon-based systems)\cite{Banholzer1992,Simon2005,Churchill2009,Balasubramanian2009,Sailer2009,Wild2012,Steger2013} to millions (in GaAs-based QDs).\cite{Hanson2007,Urbaszek2013,Kloeffel2013} In recent years, the experimental control of spin qubits in QDs has developed to a state of perfection at the single- and two-qubit level.\cite{Hanson2007,Urbaszek2013,Kloeffel2013} It is possible to initialize, manipulate,\cite{Mason2004,Petta2005,Koppens2006,Nowack2007,Koppens2008,Foletti2009,Petta2010,Bluhm2010,Pla2013,Kawakami2014,Greilich2009,Press2010,Kim2010} and read-out\cite{Elzerman2004,Morello2010,Nowack2011} spin qubits with a very high precision and to even engineer the state of the nuclear spin bath\cite{Klauser2006,Ramon2007,Baugh2007,Schuetz2014,Economou2014,Bluhm2010a,Chekhovich2010,Gullans2010,Reilly2008,Xu2009,Latta2009,Vink2009,Makhonin2011,Chekhovich2013} to increase coherence times. These remarkable experimental achievements have been accompanied by sophisticated theoretical research,\cite{Khaetskii2002,Merkulov2002,Schliemann2002,DeSousa2003,Coish2004,Shenvi2005,Coish2008,Cywinski2009a,Coish2010} which has promoted the understanding of spin dynamics in QDs with regard to the hyperfine interaction.
Therefore, we believe that these systems are ideally suited to  study predictions related to quantum equilibration.

First, we have to introduce a general theory of equilibration of the closed quantum system that fulfills all the requirements of the realization that we have in mind. This will be done on the basis of the {\it distinguishability}\cite{Short2011,Short2012} which is a measure to distinguish the actual state of a quantum system from its equilibrium state on the basis of a finite set of observables. If the values of the distinguishability are on average smaller than a given reference value $\eps$ we argue that the quantum system is $\eps$-equilibrated. In order to connect our concept of equilibration with experimentally measurable predictions, we first relate the distinguishability with the {\it weak distinguishability},\cite{Reimann2008,Short2011,GarciaPintos2015} which offers an equivalent description of equilibration under special conditions (for two-outcome observables). The time-averaged weak distinguishability (\tad), however, is capable to bound variances of expectation values from above. We have analytically derived certain bounds for the \tad, which depend on the Hamiltonian and the initial state of the quantum system. As a consequence, our analytical equilibration bounds for the \tad should directly affect the experimentally determined variance of the measurement operator. Therefore, it should be possible to modify the system at hand such that the bounds are varied and to see the difference in a direct measurement of the variance. Evidently, this is a concrete prediction of an observable consequence of quantum equilibration.

With this prediction at hand, we eventually try to better understand quantum equilibration by looking at our central spin model mentioned above. In order to calculate the \tad here, we treat very simple observables like the electron spin operator in direction parallel or perpendicular to an external magnetic field. Since we employ exact diagonalization\cite{Schliemann2002,Shenvi2005,Sarkka2008,Erbe2012,Fuchs2013} for this calculation, we are limited to a finite number of nuclear spins (up to 10). However, in state of the art QDs based on silicon or carbon host materials, such numbers of nuclear spins are within experimental reach.\cite{Banholzer1992,Simon2005,Churchill2009,Balasubramanian2009,Sailer2009,Wild2012,Steger2013} Hence, the finite size effects we analyze in this article should eventually be experimentally relevant. We find that our analytical results of bounds of the quantum equilibration describe very well the numerical simulations based on the central spin model for compatible conditions. This makes us confident that our predictions can be really seen in measurements of the spin dynamics of a confined electron spin coupled to a bath of nuclear spins.

The article is organized as follows. In Sec.~II, we explain the notion of equilibration employed in this work and introduce the (weak) distinguishability used to describe it. Subsequently, in Sec.~III, we will derive analytical results of equilibration bounds. In Sec. IV, these general results are then compared to a central spin model of an electron spin in a QD coupled to a quantum bath of nuclear spins. We conclude in Sec. V with a summary of our main results. Some derivations are presented in three Appendices.

\section{Basic concepts of equilibration}
\label{sec:definitions}
% Main part, definitions

In this section, we briefly describe known concepts of quantum equilibration for future reference.
We consider a closed quantum system whose state $\orho(t)$ evolves according to the von Neumann equation
$\dot{\oR}(t)=\frac{\I}{\hbar}\,[\oR(t),\oH]$ where $\oH$ is the $d$ dimensional Hamiltonian of the total system $\mathcal{H}$. Due to the unitary time evolution, each finite quantum system obeys a recurrence time $T_R>0$, at which the state of the system approaches within some accuracy  its initial state. However, this time does not play a role in most experiments as it scales exponentially\cite{Thirring2002} with the dimension of $\mathcal{H}$ and is almost always much larger than the age of the universe. With the commonly used and well-defined time-averaged state\cite{Short2011,Reimann2008}
\begin{equation}
\oomega := \mean{\orho}_\infty = \lim_{t\to\infty} \mean{\orho}_{t}
\label{eq:oomega}
\end{equation}
one circumvents recurrence problems. Throughout this article, $\mean{f}_t=\frac{1}{t}\int_0^t \mathrm{d}{t'}\,f(t')$ is used to denote time averages. This time-averaged state can be considered as an equilibrium state for several reasons. First, it does not evolve in time as $[\oomega, \oH] = 0$. More importantly, if the expectation value $O(t):=\tr{\oO \orho(t)}$ of any observable $\oO$ saturates at some value for long times, it can be calculated by $\mean{O}_\infty = \tr{\mean{\rho(t)}_\infty\oO}=\tr{\oomega \oO}$. In contrast to thermal states like the Gibbs state, $\oomega$ generally depends on the initial state $\orho$.

Analogously to earlier works\cite{Short2011, Reimann2008, Short2012}, we regard a quantum mechanical system to be  in equilibrium if one cannot distinguish between the state $\orho(t)$ of the full system and its equilibrium state $\oomega$ for most times by applying a  finite set of measurements $\mathcal{F}=\{ \oO_i\}$ that can be performed in an experiment.
These measurements are not restricted to subspaces of the whole Hilbert space. Hence, this definition does  not rely on the subdivision of the full quantum system into a small, measurable system  and a large, not measurable bath.

For the above notion of equilibration, it is not sufficient that the expectation value of an observable $\oO=\sum_{i}\lambda_{j}\oP_{j}$ saturates, since $\rho(t)$ and $\omega$ can still be distinguished by the (experimentally) measurable probabilities $\tr{\orho(t) \hat{P}_j}$ of its eigenvalue $\lambda_{j}$. Rather each of these time-depend probabilities has to saturate in order to guarantee indistinguishability. Considering this necessity, Short\cite{Short2011} has introduced the distinguishability 
\begin{equation}\label{eq:distinguishability}
d_\mathcal{F}(\orho(t),\oomega) = \max\limits_{\oO\in\mathcal{F}} \frac{1}{2} \sum_j \vert \tr{\oP_{j}\orho(t)}-\tr{\oP_{j}\oomega}\vert
\,.
\end{equation}
as a proper measure of distance between $\orho(t)$ and $\oomega$. Mathematically, it is closely related to the trace distance, but considers the finite number of accessible measurement operators. In contrast to the trace distance, however, this measure is not a metric, but a semi-metric since $d_\mathcal{F}(\orho(t),\oomega)=0$ is possible for $\oR(t)\neq\oomega$. This behavior is important, because it permits the desired property of equilibrated states: A sufficient condition for equilibrium is that one is not capable to distinguish the state of the system $\orho(t)$ from $\oomega$ for most times by the set $\mathcal{F}$ of measurements.

In order to account for the fact that the state of the system must be indistinguishable for most times during the time evolution, one can demand the time-average of the positive quantity $d_\mathcal{F}(\orho(t),\oomega)$ to be small.\cite{Short2011} Consequently, we regard  a system to be $\eps$-equilibrated at time $t$ if
\begin{equation}
\mean{d_\mathcal{F}(\orho(t),\oomega)}_t < \eps
\,,
\label{eq:dist_eps}
\end{equation}
where $\eps$ is a small positive constant, which we are free to choose. A reasonable choice for this constant is, for instance, the precision of the measurement devices in an experiment. Further, we call systems equilibrating in a time interval $I$ if the time-averaged distinguishability $\mean{d_\mathcal{F}(\orho,\oomega)}_t$ decreases on average within~$I$. 

\begin{figure}\centering
\includegraphics[width=0.95\linewidth]{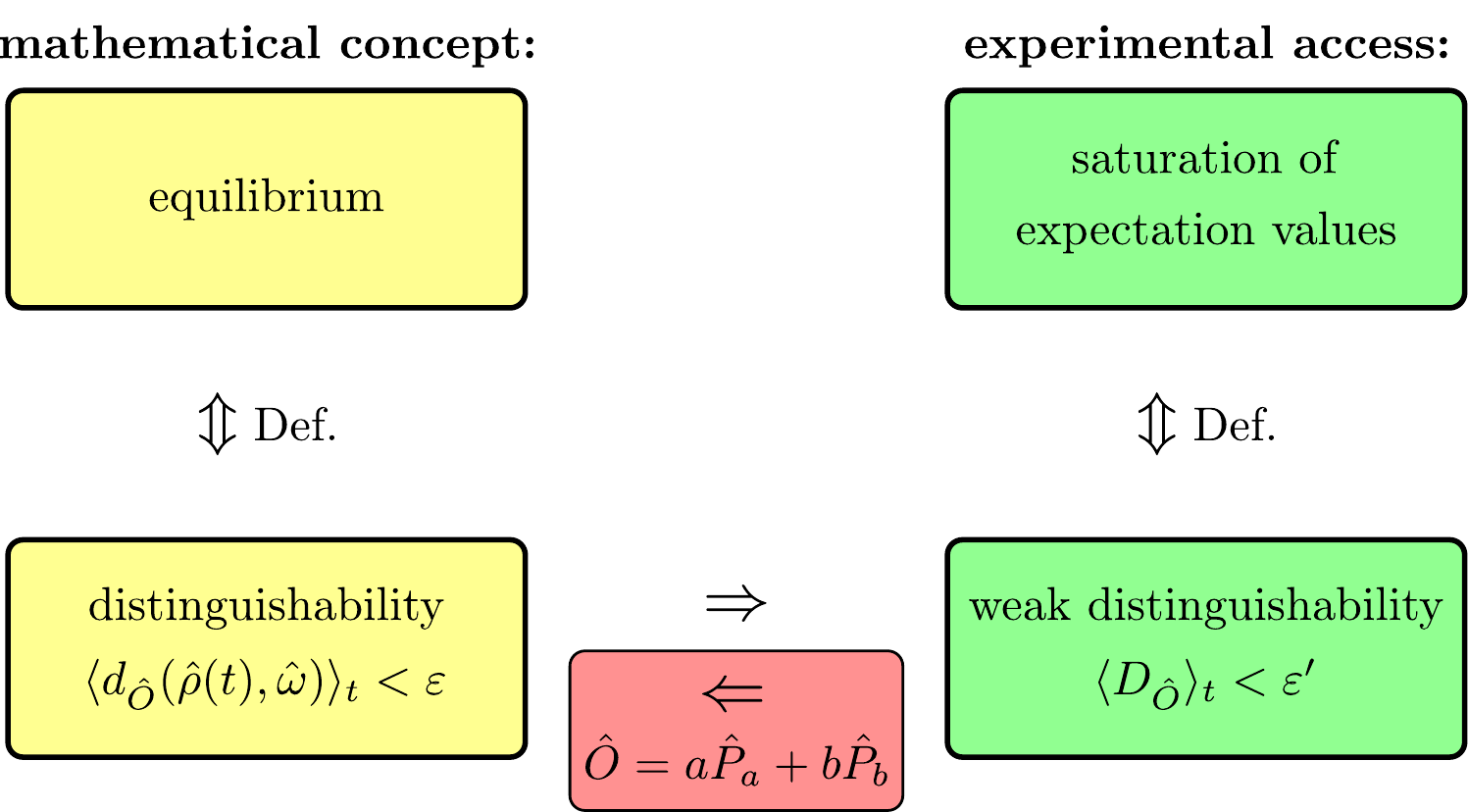}
\caption{(Color online) Connection between equilibrium, distinguishability, weak distinguishability, and expectation values. For clarity within this figure, we assume that only one observable $\oO$ is measurable, i.~e. $\mathcal{F}=\{\oO\}$. If $\oO$ is a measurement with only two possible outcomes $a\neq b$,  the weak distinguishability is equivalent to the distinguishability. Therefore, saturated expectation values are then a necessary and sufficient condition for equilibrium.}
\label{fig:equilibrium}
\end{figure}

By introducing the distinguishability and its time average in \refeq{eq:dist_eps}, we achieved a suitable mathematical definition of our concept of equilibration. However, the distinguishability cannot be measured directly in an experiment. Yet, with a slight modification of the distinguishability, one can find the so-called weak distinguishability\cite{GarciaPintos2015, Short2011,Reimann2008,Reimann2012}
\begin{equation}
\label{eq:dist}
\dist{\oO}(t) := \left(\tr{\orho(t)\oO} - \tr{\oomega\oO}\right)^2
.
\end{equation}
This quantity is unlike the distinguishability $d_\mathcal{F}(\orho(t),\oomega)$ only given by the expectation values of $\oO$\footnote{As the distinguishability can only become small if $\orho$ and $\oomega$ cannot be distinguished by any $\oO\in\mathcal{F}$, we will study the contribution a single but arbitrary observable $\oO$} with respect to both $\orho$ and $\oomega$, but does not depend on the probabilities to measure individual eigenvalues. Hence,
the weak distinguishability carries the same unit as the squared measurement operator and takes values between 0 and $4\Vert\oO\Vert^2$, where $\Vert\oO\Vert$ is the spectral norm\footnote{The spectral norm of a hermitian matrix $A$ is the eigenvalue of $A$ with the largest absolute value.} of $\oO$.
As we will show below, the long-time average of the weak distinguishability can be identified with the variances of the observable, which can be determined in an experiment. 
A small time-averaged weak distinguishability (\tad) $\mean{\dist{\oO}}_t<\eps' \ll \Vert \oO\Vert^2$ is a necessary condition for $\mean{d_{\oO}(\orho(t),\oomega)}_t<\eps\ll 1$ and, hence, according to \refeq{eq:dist_eps} for the system to be in  equilibrium. 
If two-outcome measurements $\oO = a \oP_a + b \oP_b\;(a\neq b, \oP_a + \oP_b=\tone_d)$ are considered, both quantities are even equivalent as they are then related to each other by 
\begin{equation} \label{eq:equivalence}
 \dist{\oO}(t)=(a-b)^2 [d_{\oO}(\orho(t),\oomega)]^2.
\end{equation}
In Fig.~\ref{fig:equilibrium} we summarize these dependencies and the connection to equilibrium. 
As a last property of the \tad, we show in App.~\ref{app:tawdsaturates} that the \tad
\begin{equation}
\mean{\dist{\oO}}_t = \ittad{\oO} + \delta_{\oO}(t)
\label{eq:dist_sep}
\end{equation}
is separable in a time-independent part $\ittad{\oO}$ and a time-dependent part $\delta_{\oO}(t)$, which decreases at least with $\delta_{\oO}(t)=\mathcal{O}(t^{-1})$ for $t\to\infty$. This behavior will play an important role for relating the \tad to measurable quantities in the next section.

\section{Equilibration bounds}
\label{sec:properties}

\subsection*{Weak distinguishability \textit{vs.} variance}
As argued above, the \tad is a useful quantity to describe equilibration in closed quantum systems. Moreover, it is directly related to measurable properties of the system under consideration. As we explicitly derive in App.~\ref{app:variance}, the variance $\var_{\oO}(t,\Delta t)$ of expectation values $O(t')$ in a time interval $t'\in I = [t, t + \Delta t]$ is bounded by
\begin{equation}
\label{eq:variancebound}
 \var_{\oO} (t,\Delta t) \leq \mean{\dist{\oO}}_t,
\end{equation}
where the size $\Delta t$ of the time interval $I$ needs to be sufficiently large. More precisely, $\Delta t$ must be of such a size,  that $\mean{\dist{\oO}}_t$ does not increase on average  within $I$. The above estimate even turns into an equality if $\mean{\dist{\oO}}_t$ is constant within $I$. According to \refeq{eq:dist_sep}, this is the case for each system and all observables at long times because the \tad converges.  Consequently, its infinite-time limit
\begin{align}
\label{eq:varianceinfinity}
\ittad{\oO} = \lim_{t\to\infty} \mean{\dist{\oO}}_t = \lim_{t\to\infty} \var_{\oO} (t,\Delta t) 
\end{align}
equals the variance of expectation values in any time interval $I$ at long times.  Since the time-dependent part $\delta_{\oO}(t)$ can decay much faster than $\mathcal{O}(t^{-1})$, this saturation will be already reached within finite times for many systems. Note that $\ittad{\oO}$ is the variance of expectation values of $\oO$ that does not arise from measurement errors but from fluctuations within the finite quantum system. Hence, a valid interpretation of $\ittad{\oO}$ is a measure of the capability of the system to equilibrate with respect to $\oO$. The smaller $\ittad{\oO}$ the less fluctuations of the expectation values of $\oO(t)$ around $\tr{\oomega \oO}$ are present.

\subsection*{Useful equilibration bounds at large times}

As we elaborately show  in App.~\ref{app:infinitetimeestimates}, the long-time values of the \tad can be estimated in different manners giving rise to the  bounds
\begin{align}
\ittad{\oO} \leq \ittadest{\oO}^1 &:= N_\text{G} \Vert\orho\Vert^2 \tr{\oO^2} \label{eq:firstbound}\,,\\
\ittad{\oO} \leq \ittadest{\oO}^2 &:= N_\text{G} \Vert\oO\Vert^2 \tr{\orho^2}\label{eq:secondbound}\,,\\
\ittad{\oO} \leq \ittadest{\oO}^3 &:= N_\text{G} \frac{\Vert\oO\Vert^2}{d_\mathrm{eff}}\label{eq:thirdbound}
\,.
\end{align}
Before we discuss and compare these findings, we focus on the quantities they depend on. First, in all bounds the maximum degeneracy  $N_\text{G}$ of gaps in the energy spectrum of the Hamiltonian enters, whose size is, hence, crucial for them to be of reasonable magnitude. Note that it is not sufficient to have a non-degenerate eigenvalue spectrum in order to reach $N_\text{G}=1$. \footnote{For instance, although each  eigenvalue $E_n=\hbar \omega (n+\frac{1}{2})$ of a one-dimensional harmonic oscillator is non-degenerate, the gap $\hbar\omega$ is infinite-times degenerate.}
The properties of the observable enter the equations by $\Vert\oO\Vert$ and $\tr{\oO^2}$, which are related to each other by $\Vert\oO\Vert^2 \leq \tr{\oO^2}\leq \,\Vert\oO\Vert^2\,\textrm{rank}\,\oO$.\footnote{The rank of a hermitian matrix is equals the number of its non-zero eigenvalues}
The bounds also respect the consequences of different initial states $\oR$. Explicitly, $\tr{\orho^2}$ is its purity and $\Vert\orho\Vert$ is the maximum eigenvalue of the initial state, where $\tr{\orho^2},\norm{\orho}\geq d^{-1}$. Moreover, the initial state also determines the size of the so-called effective dimension\cite{Short2011} $d_\text{eff}$, which is defined by $d_\text{eff}^{-1} = \sum_{j}(\tr{\hat{E}_j \orho})^2$ with $\hat{E}_j$ being the projector onto the eigenspace of energy $E_j$.
Vividly, $d_\text{eff}$ quantifies the dimension of the Hilbert space that is actually reached during the time evolution. It reaches values between 1 and $d$. The latter is the case for the totally mixed state $\orho=\frac{1}{d} \tone$ or for pure states like $\ket{\psi}=\frac{1}{\mysqrt{d}} \sum_j \ket{E_j}$. 

Due to the  last property of $d_\text{eff}$, the third estimate $\ittadest{\oO}^3$, which has previously been found in Ref.~\onlinecite{Short2012} with a different approach, is the most restrictive bound if pure initial states are considered. The other two estimates are useful for mixed states as both  $\Vert \orho \Vert$ and $\tr{\orho^2}$ become small if and only if the state $\orho$ is mixed.
The advantage of $\ittadest{\oO}^2$ is that the quantity $\tr{\orho^2}$ is independent of the basis whereas one needs to know all eigenstates and eigenvalues of $\oH$ in order to calculate  $d_\text{eff}$.
The bound $\ittadest{\oO}^1$ is more restrictive than $\ittadest{\oO}^2$, $\ittadest{\oO}^3$, and previously found estimates\cite{Reimann2008,Short2011}, if
\begin{equation}
d_\text{eff} \,\Vert\orho\Vert^2 \,\text{rank}\,\oO \leq 1
\,.
\end{equation}
Thus, the rank of $\oO$ should be small while the mixture of the initial state $\orho$ should be high, since $d_\text{eff}$ scales as $d$ for very mixed states while $\Vert\orho\Vert^2$ scales as $d^{-2}$. 

\subsection*{Generalization to finite times}

So far we have focused on the behavior of the \tad for long times. However, according to Eqs.~(\ref{eq:dist_sep}) and~(\ref{eq:variancebound}), we can even give estimates for finite times provided that one can bound the time-dependent part $\delta_{\oO}(t)$. As we have discussed, $\delta_{\oO}(t)$ is at least decaying as $t^{-1}$ in the long-time limit. In a recent analysis, L. P. Garc\'{i}a-Pintos and coworkers\cite{GarciaPintos2015}  have derived many interesting properties of the \tad. Among other things, the authors have bounded the time dependent part of the \tad by $\delta_{\oO}(t) \leq \frac{\lpineq}{t}$, where $\lpineq$ is a constant that dependents on $\orho, \oH$, and $\oO$. Combining this equation with our previous results, we find:  Given a system with arbitrary initial state $\orho$, Hamiltonian $\oH$ and an arbitrary observable $\oO$, we can estimate the variance of expectation values around the long-time average $\tr{\oomega\oO}$ within any time interval  $[t, t+\Delta t]$ by
\begin{equation}
\label{eq:combinedbound}
 \var_{\oO}(t,\Delta t) \leq \mean{\dist{\oO}}_t \leq  \frac{\lpineq}{t} + \min_i\ittadest{\oO}^i.
\end{equation}
The infinite-time bounds $\ittadest{\oO}^i$ are given in Eqs.~(\ref{eq:firstbound}) to~(\ref{eq:thirdbound}). 
If one of the $\ittadest{\oO}^i$ turns out to be a small number and $\oO$ is a two-outcome measurement, the system will equilibrate in the sense defined in Sec.~\ref{sec:definitions}. In that sense, $\ittadest{\oO}^i$ gives an estimate for the ability of a closed quantum system to equilibrate. Even if this concept of equilibration is not used, the above bounds still estimate the variances of observables in any closed system correctly.

\section{Application to spin models}

\subsection{Central spin model basics}
\label{sec:sub:model}
In this section, we apply the general concepts of equilibration explained above to a specific, realistic system. This allows us to show the physical significance of the above ideas for experiments. In particular, we make concrete predictions on measurable properties of an electron spin in a QD,\cite{Schliemann2003,Coish2009,Hanson2007,Urbaszek2013,Kloeffel2013} which is interacting with the nuclear spins of the host material. 

Besides this hyperfine interaction (HI) between the electron spin and the nuclear spins, we consider an external magnetic field, which is commonly used to split the Zeeman levels of the spins. In many experimental setups, further effects such as direct  interactions between nuclear spins and spin-orbit mediated effects are negligible.\cite{Hanson2007,Urbaszek2013,Fuchs2013} These interactions are, thus, not taken into account in our model. By this choice of the interactions, we construct a minimal model, which is realized by several experimental setups.\cite{Schliemann2003,Coish2009,Hanson2007,Urbaszek2013,Kloeffel2013}

Prominent examples are  devices made from group IV elements, which exhibit nuclear spin-less isotopes.
Isotopic purification of carbon or silicon allows to manipulate the number of nuclear spins present in the QD.\cite{Banholzer1992,Simon2005,Churchill2009,Balasubramanian2009,Sailer2009,Wild2012,Steger2013} This possibility allows to probe the influence of the system size on our bounds  in  Eqs.~(\ref{eq:firstbound})  to~(\ref{eq:thirdbound}).

In the following, we are especially interested in how the nuclear spins will equilibrate the electron spin. Since the observables of the electrons spin $\oS_{x,y,z}$ all have  two outcomes, the distinguishability and the weak distinguishability are equivalent according to \refeq{eq:equivalence}. The saturation of the expectation values of spin operators, hence, corresponds to the equilibration of the full system - given that they are the only accessible measurements. 

After these general considerations, let us introduce the total Hamiltonian $\oH=\oH_{\HI}+\oH_{\ZE}$ describing our model in more detail. Although our qualitative results are independent of this choice, we choose a graphene QD\cite{Trauzettel2007,Recher2010,Fuchs2012,Fuchs2013} as a reference in order to benefit from previous results.\cite{Fuchs2013} Then, the HI Hamiltonian is given by
\begin{equation}
\oH_{\HI}
=
A_{\HI}\sum_{k=1}^{K}|\phi_{k}|^2\;[\oS_{z}\oI_{z}^{k} - \frac{1}{4}(\oS_{+}\oI_{-}^{k}+\oS_{-}\oI_{+}^{k})]
\,,
\label{eq:ahi}
\end{equation}
where the energy scale of the HI\cite{Yazyev2008,Fischer2009a} is $A_{\HI}=0.6\,\mu\mathrm{eV}$ and the number of nuclear spins is $K$. We use dimensionless spin operators $\oS_{x,y,z}$, $\oI_{x,y,z}$, $\oS_{\pm}=\oS_{x}\pm\I\oS_{y}$, and $\oI_{\pm}$, analogously. The probability to find the electron at the site of the $k$-th nuclear spin is given by the absolute value of the envelope function $|\phi_{k}|^{2}\geq|\phi_{k-1}|^{2}$. The strongest HI coupling $A_{\HI}|\phi_{K}|^2$ defines the characteristic time $\tau_{\HI}=\hbar/(A_{\HI}|\phi_{K}|^2)$. Whenever we average over different initial conditions, we maintain a maximum ratio of $|\phi_{k}|^{2}/|\phi_{j}|^{2}<100$ for all $k,j$. For qualitative results, we present the results for an exemplary set of coupling constants as we have found similar results for many randomly generated sets of coupling constants.

The effect of an external magnetic field $B_{z}$ is described by the Zeeman Hamiltonian
\begin{equation}
\oH_{\ZE}
=
b\cdot  A_{\HI}|\phi_{K}|^2 \left(\oS_{z} - \gamma  \sum_{k=1}^{K}\oI_{z}^k\right)
\,,
\label{eq:aze}
\end{equation}
where $b$ is the strength of the field in units of the strongest HI coupling. Note that the nuclear spins couple only very weak to external magnetic fields compared to the electron spin: in carbon we find $\gamma = \frac{g_\text{N} \mu_N}{g \mu_\text{B}} \approx 4\times 10^{-4}$.

Besides the Hamiltonian, the time evolution of observables depends on the initial state $\oR$. In the following, we choose  product states $\oR=\oR_{\text{el}}\otimes\oR_{\text{nuc}}$, where $\oR_{\text{el}}=\ket{\psi_\text{el}}\bra{\psi_\text{el}}$ and $\oR_{\text{nuc}}$ describe the uncorelletad initial states of the electron and nuclear spins, respectively.
This assumption is plausible since the initial state of the electron spin can be experimentally well prepared in a pure (polarized) state by means of an external magnetic field\cite{Hanson2007}, using light in optically active QDs\cite{Greilich2009,Press2010,Kim2010,Urbaszek2013}, or by suitable pulse sequences in double QD setups.\cite{Mason2004,Petta2005,Koppens2006,Nowack2007,Koppens2008,Foletti2009,Petta2010,Bluhm2010,Kloeffel2013,Pla2013,Kawakami2014}  The nuclear spins, however, will on average be in an unpolarized state if no further efforts are undertaken in an experiment. Since experimentally relevant temperatures are on the order of mK to K\cite{Hanson2007,Urbaszek2013}, the thermal energy exceeds all other energy scales of the nuclear spins by far. On top of that, to follow the time evolution of the electron spin, many repetitions of the experiment are needed. Since each of these runs start with a different initial state, the nuclear spin state can be described by a totally mixed state $\oR_{\bath}=\tone/2^K$ on average. However, the nuclear spin state can be also manipulated by means of dynamical nuclear polarization\cite{Ramon2007,Baugh2007,Chekhovich2010,Gullans2010,Economou2014} and state narrowing,\cite{Klauser2006,Reilly2008,Xu2009,Latta2009,Vink2009,Bluhm2010a} which allow to significantly polarize the nuclear spins and to change the composition of the initial state of the nuclear spins.
\footnote{Optical polarizations in the range of $\overline{p}\approx0.6$ are now routinely achieved\cite{Chekhovich2013}. In electrically controlled QDs, realized polarizations typically are in the range of percent, but also in these systems a polarization of $\overline{p}\approx0.4$ has been reported.\cite{Chekhovich2012}}
Motivated by these experimental possibilities, we also investigate the effect of polarized initial states of the nuclear spins by using a Gaussian distribution of states.

With both, the Hamiltonian and the initial state given, the time evolution of the density matrix and, hence, of every observable in the system can be calculated by exact diagonalization\cite{Schliemann2002,Shenvi2005,Sarkka2008,Erbe2012,Fuchs2013},  which is performed using the {\small EIGEN}\cite{Guennebaud2010} package for {\small C++}.

\subsection{Spin dynamics}
\label{sec:sub:results}

Once the time evolution of an observable is known, its variance, the weak distinguishability and the \tad defined in Sec. \ref{sec:definitions} are readily calculated. This enables us to demonstrate, that the \tad indeed bounds the variances of an observable.
\begin{figure}
\centering
\includegraphics[width=0.98\linewidth]{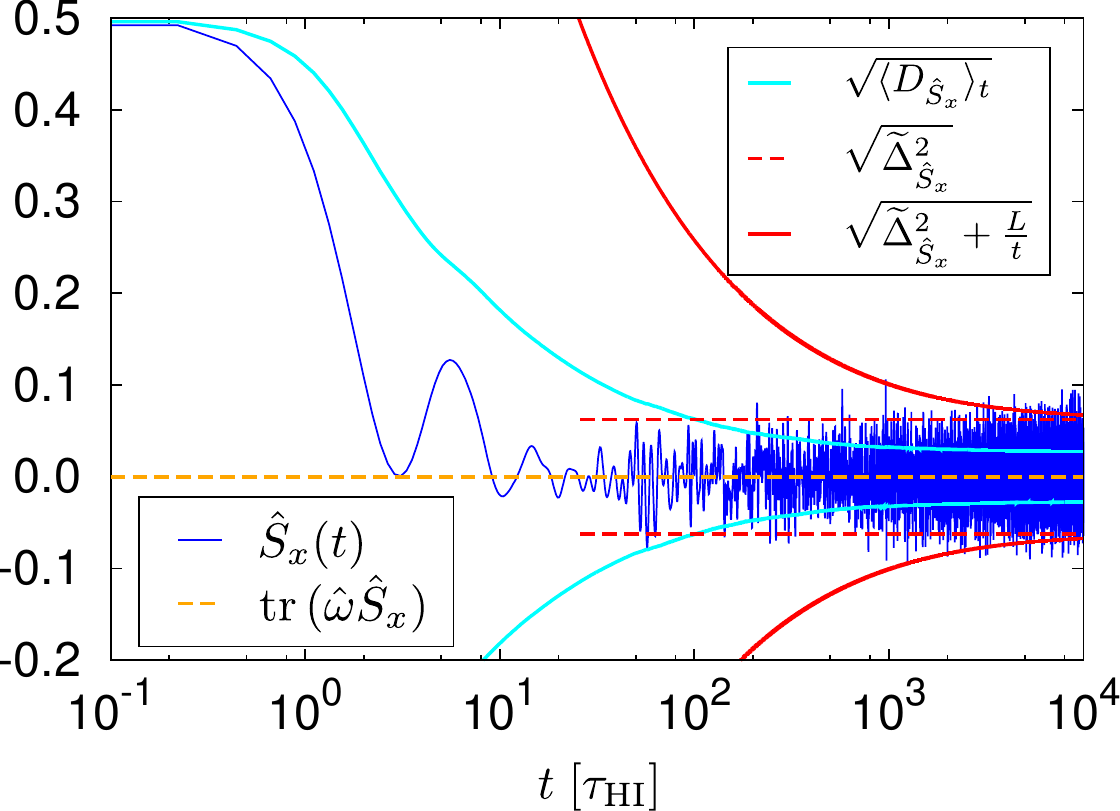}
\caption{(Color online) The time evolution of the electron spin component $S_{x}(t)$ and its long-time average $\tr{\oomega\oS_{x}}$. We use a magnetic field $b=\frac{1}{4}$ and $K=6$ nuclear spins with random coupling constants. Initially, the electron spin is maximally polarized in $x$-direction while the nuclear spins are in the totally mixed state. The square root of the \tad $\mysqrt{\av{\dist{\oS_{x}}}_{t}}$ bounds the standard deviation of $S_{x}(t)$ for all times and converges to it for large times. For comparison, we plotted the analytically derived bounds given in \refeq{eq:combinedbound} and \refeq{eq:secondbound}. Note that all quantities are dimensionless.}
\label{fig:variance}
\end{figure}

As an example, we show the evolution of $S_x(t)$ in Fig.~\ref{fig:variance}. At times $t\sim \tau_\text{HI}$, the initially polarized electron spin begins to oscillate with decreasing amplitude around its long-time average $\av{S_{x}}_{\infty}\approx0$. The square root of the weak distinguishability $\mysqrt{\av{\dist{\oS_{x}}}_{t}}$ bounds the standard deviation of $S_x(t)$ as predicted at all times. At large times, the \tad $\av{\dist{\oS_{x}}}_{t}$ saturates to a finite value whose size corresponds to the quantum fluctuations in our finite model. As explained above, the \tad in turn can be bounded itself by the analytical expression given in \refeq{eq:combinedbound}. For finite times, this bound decays with $\mathcal{O}(t^{-1})$, while it saturates at $\ittadest{\oS_{x}}^{2}$ given in \refeq{eq:secondbound} for large times. For the parameters chosen in Fig.~\ref{fig:variance}, we find $\ittadest{\oS_{x}}^{2}=2^{-(K+1)}$. \footnote{We use \refeq{eq:secondbound} with $N_{G}=1$, $\tr{\oR^{2}}=\frac{d}{2}(\frac{2}{d})^{2}$, $\Vert\oS_{x}\Vert^{2}=\frac{1}{4}$, and $d=2^{K}$}
 Remarkably, already for $K=6$ nuclear spins, this long-time estimate yields a very sharp upper bound on the standard deviation of fluctuations of the signal.

As explained above, the properties of the \tad and its bounds depend on the Hamiltonian of the system. Thus, one should test how different Hamiltonians alter the equilibration.
\begin{figure}
\centering
\includegraphics[width=0.98\linewidth]{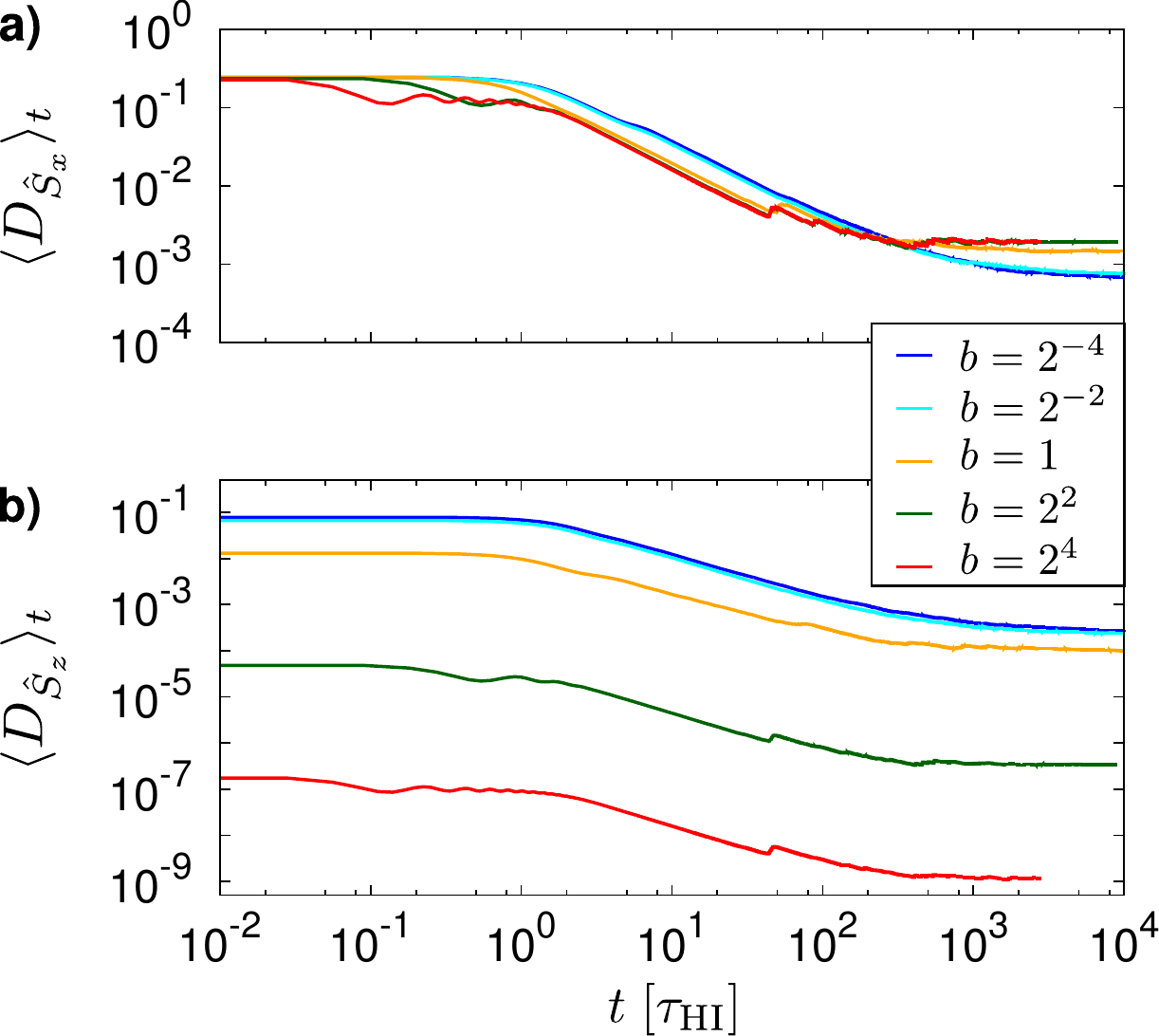}
\caption{(Color online) The \tad as a function of time for different external magnetic fields $b$ for \textbf{a)} $\hat{S}_x$ and \textbf{b)} $\hat{S}_z$. The electron spin is initially polarized in $x$- and $z$-direction, respectively. Apart from that, the same parameters as in Fig.~\ref{fig:variance} have been used. \textbf{a)} All \tad{}~s start with the same value and show oscillations on a timescale $\tau_{\ZE}\propto b$. The actual equilibration starts at $\tau_{\HI}$. \textbf{b)} The \tad{}s start with a different value for different $b$. The equilibration due to the HI also starts at $\tau_{\HI}$.}
\label{fig:magnetic_field}
\end{figure}
In a QD, the easiest way to change the Hamiltonian is to modify the external magnetic field. By varying $b$ over approximately two orders of magnitude, we sweep from a situation in which the electron spin couples most strongly to the nuclear spins to a scenario where the Zeeman coupling is dominant. In Fig.~\ref{fig:magnetic_field}, we compare the \tad{}s of $\hat{S}_x$ and $\hat{S}_z$. For both spin components, we observe that equilibration sets in approximately at time $\tau_\HI$ and reduces the initial values of the \tad{}s roughly by two orders of magnitude for all values of $b$. As we discuss later, the size of this reduction depends on the number of nuclear spins. In fact, even high values of $b$ cause only Lamor oscillations of $\hat{S}_{x}$ at small times $\tau_{\ZE}\propto b$, but do not change the overall equilibration behavior. Besides this, the only effect of large magnetic fields is a reduced initial value of the \tad for $\hat{S}_z$. This can be understood as follows. As the electron spin is initially fully polarized parallel to a strong magnetic field, its initial state is almost fully preserved, since the flip-flop terms of the HI are suppressed due to the large Zeeman splitting of the electron spin states. In other words, the electron is initially approximately in an eigenstate of the total Hamiltonian for strong external magnetic fields. Hence, $\orho$ is already initially close to $\oomega$ and, as a consequence, indistinguishable from $\oomega$ by means of $\oS_{z}$.

Similar effects can be found for polarized states, which we approximate by a Gaussian distribution of states characterized by a mean polarization $\bar{p}$ and a standard deviation $\sigma_{p}$. First, we have fixed $\sigma_{p}=0.3$ and varied the mean polarization between $p=0$ and $p=0.75$ for a system containing $K=6$ nuclear spins. Since the initial states approach eigenstates of the total Hamiltonian for increasing polarization, we observe a decreasing initial distinguishability $\dist{\oS_{z}}(0)$. Due to the HI, the \tad $\av{\dist{\oS_{z}}}_{\infty}$ saturates again around a value, which is about two orders of magnitude smaller than its initial size. For larger polarizations this reduction becomes smaller, since the HI spin flip-flops become less effective. These findings are consistent with a smaller effective dimension $d_{\mathrm{eff}}$ of polarized initial states in \refeq{eq:thirdbound}. Analogous simulations with standard deviations in an interval $0.15<\sigma_{p}<0.75$ show no significant differences to these observations.

\subsection{Size dependence of the bath}
\label{sec:sub:quantumbath}
\begin{figure}
\centering
\includegraphics[width=0.98\linewidth]{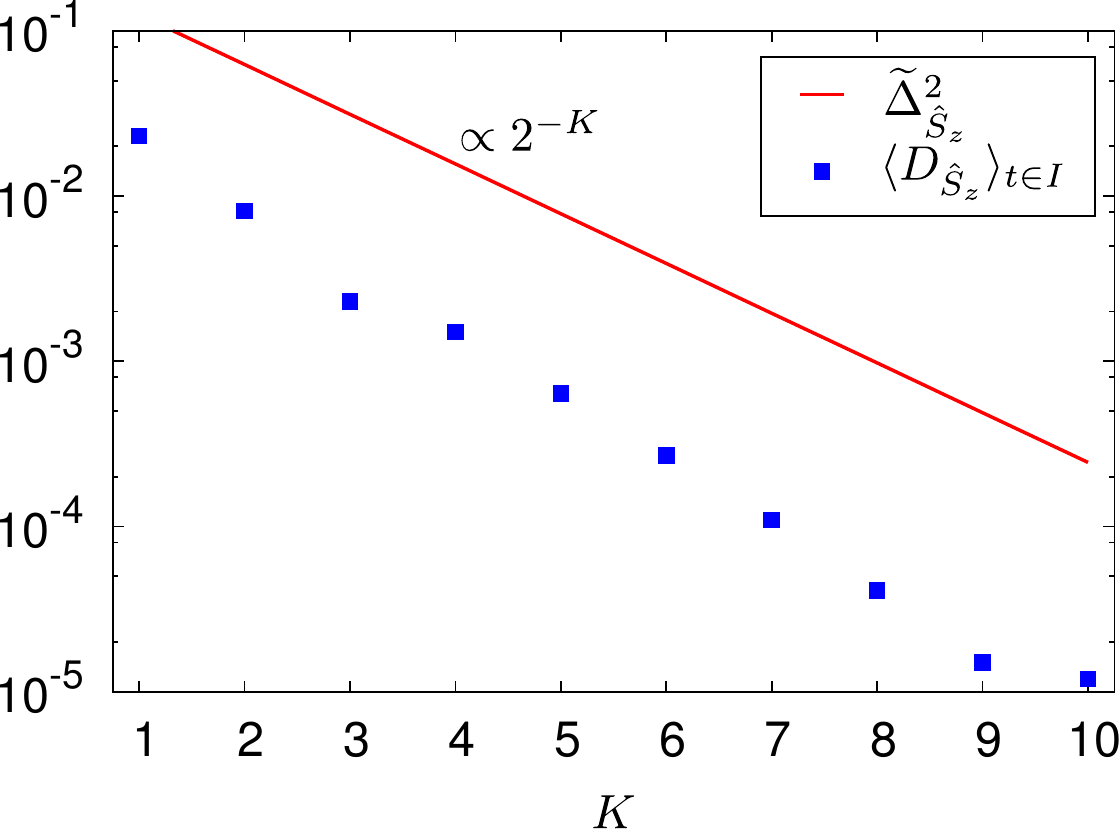}
\caption{Color online) Dependence of the long-time \tad for the observable $\oS_{z}$ on the number of nuclear spins $K$. We approximate $\av{\dist{\oS_z}}_\infty$ by $\av{\dist{\oS_z}}_{t\in I}$, where $I=[\tau,\tau+\Delta\tau]$ is a time interval with $\tau \gg\Delta\tau\gg\tau_\text{HI}$. The numerical data is compared to the analytical bound $\ittadest{\oS_{z}}^{2}$ (solid line) given in \refeq{eq:secondbound}. The fit (dashed line) suggests a $\av{\dist{\oS_{z}}}_{\infty} \sim 2^{-1.2 K}$ dependence. We average over at least 100 (40 for $K=10$) sets of random coupling constants $\abs{\phi_k}^2$ and show the mean value with the standard deviation as blue points. Besides a magnetic field of $b=0.05$ we use the totally mixed state for nuclear spins but polarize the electron spin in $z$ direction.}
\label{fig:K_dependence}
\end{figure}

We finally want to address the question how many nuclear spins are required in order to treat them as a bath. By adding more and more nuclear spins, no sudden change is observed but the fluctuation of spin components of the electron decrease exponentially with the number of nuclear spins, cf. Fig.~\ref{fig:K_dependence}.

The numerically obtained values of the long-time \tad are about one order of magnitude smaller than the presented bound $\ittadest{\oS_{z}}^{2}$. Considering the estimates made and the generality of the bound, this is still a
fairly good result. Fig.~\ref{fig:K_dependence} also suggest that quantum fluctuations may decrease even faster  with increasing system size than our analytic bounds require. Note that this $K$-dependence of the equilibration properties is
not limited to mixed states only. Reconsidering previously obtained data\cite{Fuchs2013}, we have calculated the effective dimension
for randomly chosen pure initial states of the nuclear spins. For these states, it scales approximately
with $d_{\mathrm{eff}}\sim d/2 = 2^{K}$. According to $\ittadest{\oS_{z}}^{3}\propto d_{\mathrm{eff}}^{-1}=2^{-K}$ given in \refeq{eq:thirdbound}, this dependence also gives rise to an exponential decay of $\av{\dist{\oS_{z}}}_{\infty}$, which is confirmed by our  numerics.\cite{Fuchs2013}  As discussed by Reimann\cite{Reimann2008}, the effective dimension of almost all states grow exponentially with the size of the system. Hence, such a decay is a rather generic result, which can be understood as follows. If we add a nuclear spin to the system, we double both the size of the Hilbert space and the number of energies driving the dynamics of the electron spin, which finally leads to the observed reduction of fluctuations.

We can indeed generalize these findings to other quantum systems that differ from our model, e.g. a central spin model with isotropic hyperfine interaction or even topologically different models like spin chains. 
Given that the effective dimension $d_\text{eff}\sim d $ scales exponentially with the bath size, either due to totally mixed bath states or due to randomly chosen pure initial states,\cite{Reimann2008} we can use the bound in Eq.~(\ref{eq:thirdbound}) to deduce the following statement. The number $N$ of bath spins that are sufficient to saturate an electron spin in some arbitrary quantum model increases only logarithmic with the inverse resolution $1/r$ of the measurement:
\begin{equation}
 N \geq \log_2\left(\frac{N_G \Vert O \Vert^2}{c}\cdot\frac{1}{r^2}\right)
\label{eq:bathsizedependence},
\end{equation}
where $c=d_\text{eff}/d\lesssim 1$. For a resolution $r=0.01\,\hbar$ and an initial state far away from an energy eigenstate ($c\to1$), the electron spin components equilibrate in any quantum model with non-degenerate gaps ($N_G=1$) if the electron is coupled to more than 11 bath spins.
As our model demonstrates, even less bath spins $N\approx 7$ are capable of equilibrating the electron spin components below this resolution in experimentally relevant scenarios.

\section{Summary and discussion}
In summary, we have shown how a general theory on equilibration can be applied to a realistic closed quantum system. We have introduced a specific understanding of equilibration relevant for our system under consideration and analyzed its properties by analytical calculations. Afterwards, we have applied this concept to a model of electron and nuclear spins in a solid state QD, which we have investigated by numerical simulations.

A system is assumed to be in equilibrium, if an observer cannot distinguish, for most times, between the actual state of the system and its equilibrium state using a finite set of measurements. Notably, two observers with different measurement sets could come to different conclusions. This equilibrium state is not necessarily a thermal state since it can, for instance, depend on the initial state of the system. The distinguishability between the state of the system and its equilibrium state can be quantified by a suitable ``measure of distance''. In this article, we consider the so-called weak distinguishability which vanishes whenever the system is in equilibrium. For two-outcome measurements, we have been able to show that a saturation of the corresponding expectation values is equivalent to equilibration. Furthermore, we have demonstrated how the variance of a time-dependent observable can be bounded by this weak distinguishability, which has allowed us to connect this abstract mathematical function to an experimentally measurable quantity. We have also derived three different bounds for the time-averaged weak distinguishability and thereby recovered one previously known bound by means of a new method.\cite{Short2011} We have therefore been able to predict upper limits to the size of fluctuations in small closed quantum systems. 

Applying our analytical results to a QD setup in which an electron spin is coupled to nuclear spins of the host material through the hyperfine interaction enables us to make precise predictions. Since this spin system is typically well isolated from its environment, QDs can be considered as a closed quantum system for sufficiently short time scales.  We have simulated the time evolution of the total spin system and have analyzed its dependence on experimentally accessible parameters such as the strength of an external magnetic field and the  polarization of the initial state of the nuclear spins. Intriguingly, we have discovered cases in which strong magnetic fields do not prevent the electron spin from equilibration, while a polarized bath always diminishes the equilibration capability. Finally, we have also investigated the importance of the number of nuclear spins on equilibration properties. We show both analytically and numerically that very small amounts of bath spins are sufficient to fully equilibrate the electron spin in our model. The analytical results even hold for a wider class of spin models, and, thus are not limited to our specific model.

\section{Aknowledgements}
We would like to thank L.~P.~Garc\'{\i}a-Pintos for many stimulating discussions and for sharing with us the unpublished manuscript of Ref.~\onlinecite{GarciaPintos2015}. Furthermore, we acknowledge interesting conversations with N.~Linden, P.~Reimann, A.~J.~Short, and S.~Wehner. Financial support has been provided by the Deutsche Forschungsgemeinschaft (DFG) through the priority program SSP 1449 and the research grant TR950/8-1.

\appendix

% Appendix

\section{Saturation of variances}
\label{app:tawdsaturates}
In this Appendix, we show that one can separate  $\mean{D_{\oO}}_t=\delta_{\oO}(t) + \Delta_{\oO}$ in a time dependent part $\delta_{\oO}(t)$ that vanishes at large times and a time independent part $\Delta_{\oO}$. To do so, we follow a previous analysis\cite{GarciaPintos2015} and use the fact that the matrix elements $\omega_{ij}=\bra{E_i}\oomega\ket{E_j}$ of $\oomega$  in energy space are given by
\begin{equation}
\label{eq:omegahandy}
 \omega_{ij} = \begin{cases} \rho_{ij} & E_i = E_j\\ 0 & \text{else}\end{cases}\,,
\end{equation}
where $E_i$ is the energy of the $i$-th eigenvector $\ket{E_i}$ of $\oH$. Now, we can rewrite the \tad by
\begin{align}
 &\mean{\dist{\oO}}_t = \frac{1}{t} \int_0^{\mathrlap{t}} \tot{t'} \left(\tr{\orho(t)\oO}-\tr{\oomega \oO}\right)^2\\
&= \frac{1}{t} \int_0^{\mathrlap{t}}\tot{t'} \abs{\sum_{n,m}(\rho_{nm}e^{-\I\hbar^{-1}(E_n {-} E_m)t'} -\omega_{nm}) O_{mn}}^2 \nonumber \\
&=  \frac{1}{t} \int_0^{\mathrlap{t}}\tot{t'} \abs{\sum_{n,m}(\rho_{nm} {-} \omega_{nm}) e^{-\I\hbar^{-1}(E_n {-} E_m)t'}O_{mn}}^{\mathrlap{2}}. \nonumber
\end{align}
The last step is possible because for all $n,m$ with $\omega_{nm}\neq 0$ follows $E_n = E_m$ (see Eq.~(\ref{eq:omegahandy})). Therefore, $\omega_{nm}e^{-\frac{\I}{\hbar}(E_n {-} E_m)t'}$ is time independent.  We define $v_\alpha = (\rho_{nm}-\omega_{nm})O_{mn}$ and gaps $G_\alpha = E_n - E_m$. Note that $\alpha$ is an abbreviation for a double index, running over all $d^2$ gaps. We then find
\begin{align}
&\mean{\dist{\oO}}_t  = \frac{1}{t} \int_0^t {}\tot{t'} \abs{\sum_\alpha v_\alpha e^{-\I\hbar^{-1} G_\alpha t'}}^2\\
&= \frac{1}{t} \int_0^t{}\tot{t'} \sum_{\alpha,\beta} v_\alpha v_\beta^* e^{-\I\hbar^{-1}(G_\alpha - G_\beta)t'} \nonumber \\
&= \frac{1}{t} \int_0^t {}\tot{t'}\Big( \sum_{\substack{\alpha,\beta\\G_\alpha = G_\beta}} + \sum_{\substack{\alpha,\beta\\G_\alpha \neq G_\beta}}\Big) v_\alpha v_\beta^* e^{-\I\hbar^{-1}(G_\alpha - G_\beta)t'} \nonumber \\
&= \underbrace{\frac{1}{t}  \sum_{\mathclap{\substack{\alpha,\beta\\G_\alpha \neq G_\beta}}} v_\alpha v_\beta^* \int_0^t {}\tot{t'} e^{-\I\hbar^{-1}(G_\alpha - G_\beta)t'}}_{=\delta_{\oO}(t)} + \underbrace{\;\;\;\;\sum_{\mathclap{\substack{\alpha,\beta\\G_\alpha = G_\beta\neq 0}}}  v_\alpha v_\beta^*}_{=\ittad{\oO}}.
\label{eq:app:dist}
\end{align}
We can exclude the cases $G_\alpha=G_\beta=0$ in the second term because $G_\alpha=0$ implies $v_\alpha=0$.
Note that $\delta_{\oO}(t)$ vanishes at least with $\frac{1}{t}$ in limit of infinite times because the sum in $\delta_{\oO}(t)$ is upper-bounded by $\sum_{\{\alpha,\beta\,\vert\, G_\alpha\neq G_\beta\}} \vert v_\alpha v_\beta^*\vert$. Hence,
\begin{align}
 \lim_{t\to\infty} \mean{\dist{\oO}}_t = \ittad{\oO}.
\end{align}
%Using the concept of equilibration we defined in Sec.~\ref{sec:definitions}, each system equilibrates with respect to observables with two-outcome states. This is because %$\mean{\dist{\oO}}_t$ is saturating at $\ittad{\oO}$ such that the system can neither be called approaching to an equilibrium nor veering away from an equilibrium. This looks intuitively %strange e.~g. for a single spin rotating in a constant magnetic field. However, the quality of equilibration can be quantified by $\ittad{\oO}$, which is indeed quite large for this %scenario.

\section{Relation between distinguishability and variance}

\label{app:variance}
Defining $s(t):= \frac{\tot{}}{\tot{\ln t}} \ln\mean{\dist{\oO}}_t$, one can easily rewrite the definition of $\mean{\dist{\oO}}_t$ (see  Eq.~(\ref{eq:dist})) by
\begin{equation}
 \dist{\oO}(t) = [s(t) + 1] \mean{\dist{\oO}}_t.
\end{equation}
An average over the time interval $I=[t, t + \Delta t]$ yields
\begin{align}
 \int_I \tot{t'} \dist{\oO}(t') &\leq \max_{t'\in I} \mean{\dist{O}}_{t'} \cdot  \int_I \tot{t'} [s(t')+1].
\end{align}
 With $\dist{\oO}(t) = (\tr{\orho(t)\oO} - \tr{\oomega\oO})^2$, the left hand side of the latter equation represents the variance of expectation values of $\oO$ around the value $\tr{\oomega\oO}$ within the time interval $I$. Defining $\bar{s}(t)$ to be the average slope $s(t)$ within $I$, we derive
\begin{equation}
 \var_{\oO}(t,\Delta t) \leq [\bar{s}(t) + 1] \mean{\dist{\oO}}_t,
\end{equation}
where we assume that $\max_{t'\in I}\mean{\dist{\oO}}_{t'} = \mean{\dist{\oO}}_t$. This assumption is correct if the system is on average approaching its equilibrium state. The value of $\bar{s}(t)$ is then negative, however, $s(t) \geq -1$ holds strictly.
 This  follows from both the semi-positive values of  $\dist{\oO}(t)$ and the $\frac{1}{t}$ in the definition of $\mean{\dist{\oO}}_t$. Therefore, we prove that
\begin{equation}
 \var_{\oO}(t,\Delta t) \leq \mean{\dist{\oO}}_t.
\end{equation}
The latter bound holds for all systems that approach equilibrium in the sense defined above. If a system is already equilibrated, the \tad $\mean{D_{\oO}}_t$ is no longer decreasing such that $\bar{s}(t)=0$. Note that in this limit, the estimate for the variance becomes exact. This is also the case at large times, where $\mean{\dist{\oO}}_t$ of each system saturates as we explain above in  App.~\ref{app:tawdsaturates}.

\section{Infinite time estimates}
\label{app:infinitetimeestimates}

 In the following, we show how to estimate $\ittad{\oO}$ using only basic information about the system. For this purpose, we start with the long-time limit of \refeq{eq:app:dist}
\begin{align}
 \ittad{\oO} &= \sum_{\mathclap{\substack{\alpha,\beta\\ G_\alpha = G_\beta\neq 0}}} v_\alpha v_\beta^* = \sum_j \sum_{a,b}^{n_j} v^j_a {v^j_b}^*, \label{eq:beforebox}
\end{align}
where the sum in the last step is symmetrized by defining a parameter $j$ to run over all distinct values of energy gaps, while $a$ and $b$ run over all $n_j$ gaps of size $G_j$. Therefore, $v^j_a$ belongs to the $a$-th gap of size $G_j\neq0$. We  estimate the symmetric double sum by
\begin{align}
 \sum_{i,j}^N x_i x_j^* &= \sum_i^N{\abs{x_i}^2} + \frac{1}{2} \sum_{i\neq j}^N \left( x_i x_j^* + x_j x_i^*\right) \nonumber \\
&\leq \sum_i^N \abs{x_i}^2 + \frac{1}{2} \sum_{i\neq j}^N \left(\abs{x_i}^2+\abs{x_j}^2\right) \nonumber \\
&= N \sum_{i}^N \abs{x_i}^2,
\end{align}
where $\left\{x_i\right\}$ is a set of $N$ arbitrary complex numbers. Applying this relation to Eq.~(\ref{eq:beforebox}), we obtain
\begin{align}
 \ittad{\oO}
&\leq \sum_j n_j  \sum_{a}^{n_j} \abs{v^j_a}^2,
\end{align}
which even is an equality as long as all gaps are not degenerate, \ie\,$n_j=1\;\forall j$. With the maximum degeneracy of energy gaps $N_\text{G}=\max_j{n_j}$, we find
\begin{align}
 \ittad{\oO} \leq N_\text{G} \sum_j \sum_a^{n_j} \abs{v^j_a}^2,
\end{align}
where both sums combined run over all $d^2$ gaps in the energy spectrum. In the previous notation this reads
\begin{align}%
 \ittadest{\oO} \leq N_\text{G} \sum_\alpha \abs{v_\alpha}^2.%
\end{align}%
We now insert the definition of $v_\alpha$ and use $\abs{\rho_{nm}-\omega_{nm}}\leq\abs{\rho_{nm}}$, which follows from Eq.~(\ref{eq:omegahandy}). Thus, we find
\begin{align}
\label{eq:beforesplitting}
 \ittadest{\oO} &\leq N_\text{G} \sum_{n,m}\abs{\rho_{nm} O_{mn}}^2.
\end{align}

\subsection*{First and Second Estimate}
For the first estimate $\ittadest{\oO}^1$, we use  that $\abs{\rho_{nm}O_{mn}}=\abs{\rho_{nm}}\abs{O_{mn}}$ and $\abs{\rho_{nm}} \leq \Vert\orho\Vert$, where $\Vert\orho\Vert$ is the spectral norm of $\orho$. Using this and $\oO=\oO^\dagger$, this yields
\begin{align}
 \ittad{\oO} &\leq N_\text{G} \Vert\orho\Vert^2 \sum_{nm} O_{nm} O_{nm}^* \nonumber \\
&  = N_\text{G} \abs{\abs{\orho}}^2 \tr{\oO^2} \nonumber \\
&=: \ittadest{\oO}^1
\end{align}
The same steps but $\orho$ and $\oO$ interchanged yields
\begin{align}
 \ittad{\oO}\leq \ittadest{\oO}^2 := N_\text{G} \Vert\oO\Vert^2\tr{\orho^2}.
\end{align}

\subsection*{Third Estimate}
We derive  estimate $\ittadest{\oO}^3$ by following an approach along the lines of Ref.~\onlinecite{Short2012}. We can then estimate Eq.~(\ref{eq:beforesplitting}) by
\begin{align}
\ittad{\oO} &\leq N_\text{G} \sum_{n,m} \rho_{nn} \rho_{mm} O_{mn} O_{nm} \nonumber \\
&= N_\text{G} \tr{\orho_\text{diag} \oO \orho_\text{diag} \oO},
\end{align}
where $\abs{\rho_{nm}}^2\leq \rho_{nn} \rho_{mm}$ because $\orho$ is positive and $(\orho_\text{diag})_{nm}=\rho_{nm}\delta_{nm}$. With the Cauchy-Schwarz inequality and $\tr{AB}\leq \Vert A\Vert \tr{B}$ for positive $A$ and $B$ follows
\begin{align}
 \ittad{\oO} &\leq N_\text{G} \tr{\oO^2\orho_\text{diag}^2} \nonumber \\
 &\leq N_\text{G}\Vert\oO\Vert^2 \tr{\orho_\text{diag}^2} \nonumber \\
&\leq N_\text{G}\Vert\oO\Vert^2\tr{\oomega^2} \nonumber \\
&= N_\text{G} \frac{\Vert\oO\Vert^2 }{d_\text{eff}}\\
&=: \ittadest{\oO}^3
\end{align}
This bound has previously been obtained in Ref.~\onlinecite{Short2012} on the basis of an analysis with pure initial states that have been expanded to mixed states afterwards.

% \bibliographystyle{apsrev4-1.bst}
% \bibliography{library.bib}


\end{document}